\documentclass{proc}
\usepackage{amsfonts}
\usepackage{amsmath}
	\usepackage{graphicx}
  \graphicspath{{/Figs}}
  \DeclareGraphicsExtensions{.eps}
  
 \usepackage{color}
 \usepackage{hyperref}
 \usepackage{breakurl}

\usepackage{flushend}
\usepackage{tabulary}
\usepackage{tabularx}
\usepackage{mathptmx}
\begin{document}

\title{What do they know about me?\\Contents and Concerns of Online Behavioral Profiles}
 \author{Ashwini Rao, Florian Schaub, Norman Sadeh \\
 Carnegie Mellon University\\
 arao@cmu.edu, fschaub@cmu.edu, sadeh@cs.cmu.edu
}

\maketitle


\begin{abstract}
Data aggregators collect large amount of information about individual users and create detailed online behavioral profiles of individuals. Behavioral profiles benefit users by improving products and services. However, they have also raised concerns regarding user privacy, transparency of collection practices and accuracy of data in the profiles. To improve transparency, some companies are allowing users to access their behavioral profiles. In this work, we investigated behavioral profiles of users by utilizing these access mechanisms. Using in-person interviews (n=8), we analyzed the data shown in the profiles, elicited user concerns, and estimated accuracy of profiles. We confirmed our interview findings via an online survey (n=100). To assess the claim of improving transparency, we compared data shown in profiles with the data that companies have about users. More than 70\% of the participants expressed concerns about collection of sensitive data such as credit and health information, level of detail and how their data may be used. We found a large gap between the data shown in profiles and the data possessed by companies. A large number of profiles were inaccurate with as much as 80\% inaccuracy. We discuss implications for public policy management.   
\end{abstract}

\section{Introduction}
The online services landscape is driven by a data economy in which data aggregators and service providers trade user information on data marketplaces~\cite{ftc2014broker,bluekai2014market}. As part of the data economy, data aggregators and service providers collect extensive amount of data about individuals from multiple sources, including public, online and offline sources~\cite{ftc2014broker}. By combining information from multiple sources, they create behavioral profiles of individuals. The data and profiles may be used for purposes such as personalization, risk mitigation products, people search and targeted advertising~\cite{ftc2014broker}. The data economy benefits users by providing better products and services. It also sustains many free services such as search and social networking. However, the data economy also raises privacy concerns. For example, studies have found that users have privacy concerns when behavioral profiles are used for advertising~\cite{ur2012smart, mcdonald2010beliefs, truste2011oba, agarwal2013not}. Using profile data for risk mitigation services such as background checks raises concerns about accuracy of data~\cite{ftc2014broker}.

In this work, we investigated online behavioral profiles of users. To study behavioral profiles, we used access mechanisms provided by companies that allow users to look at data in their profiles~\cite{bluekai2014registry,google2014adsetting,yahoo2014adinterest}. Companies have started providing access to improve transparency of their data collection practices. First, we analyzed the types of data found in actual profiles. A prior report by the Federal Trade Commission investigated the types of data that companies may potentially use to build behavioral profiles. However, it did not look at contents of actual profiles~\cite{ftc2014broker}. We compared the data shown in profiles with the data companies claim to possess about users. This allowed us to evaluate the claim of increasing transparency by providing access to profiles. 

Second, we studied user concerns and surprises regarding data in their behavioral profiles. Prior studies have focused on user concerns and perceptions regarding use of behavioral profiles for advertising~\cite{ur2012smart,agarwal2013not}. We focus on user privacy concerns regarding actual contents of behavioral profiles. Our approach of using user's own behavioral profile for eliciting concerns and surprises leads to a more contextualized and nuanced understanding of user concerns regarding online behavioral profiles. Further, we were able to estimate the accuracy of data in user profiles. Our study also provides insight into usability of access mechanisms.
 
To understand contents and concerns of behavioral profiles, we first conducted semi-structured interviews in which we asked our participants (n=8) to look at their own profiles. We elicited their surprises and concerns regarding the data in their profiles, and documented the types of data found in their profiles. Subsequently, we conducted an online survey (n=100) to confirm the identified user concerns with a more diverse audience. We investigated the types of data that companies possess about users by surveying documents published by data aggregators and service providers.

We organize the rest of the paper as follows. In Section~\ref{sec:profiles}, we provide background information on the data economy and access mechanisms provided by companies. We discuss related work in Section~\ref{sec:related}. In Section~\ref{sec:methodology}, we provide an overview of our study including interview details. In Section~\ref{sec:contents}, we present our findings regarding contents of actual behavioral profiles and contrast it with data possessed by companies. In Section~\ref{sec:concern}, we discuss user concerns and surprises. In Section~\ref{sec:survey}, we provide details of our online survey and its results.  Lastly, in Section~\ref{sec:discussion}, we discuss the insights gained from our study and conclude with a discussion of implications for public policy and future research.    

\section{Background}
\label{sec:profiles}
To provide necessary background for the rest of the paper, we briefly discuss the data economy and access mechanisms.

\begin{figure}
\centering
\includegraphics[trim=0 0 0 0,clip,width=2.5in]{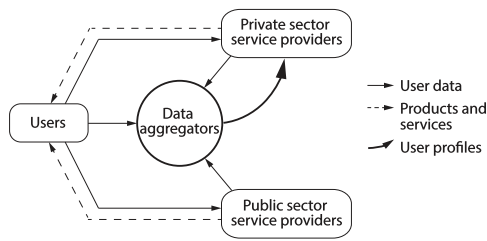}
\caption{A conceptual model of the data economy}
\label{fig:data_economy}
\end{figure}

\paragraph{The Data Economy:}
In Fig.~\ref{fig:data_economy}, we show a simple conceptual model of the data economy that highlights the role of data aggregators. Users provide their personal data to public and private sector service providers when they receive products and services from these service providers. We group all entities such as websites, offline stores, advertisers and marketers under the umbrella of private sector service providers. Data aggregators collect different types of user data available from service providers and also via direct engagement with users. Public sources of information include census data, voter registration databases, occupation data from state license boards, bankruptcy records, county deed and tax assessor records, and Yellow-pages directories~\cite{experianmarket2013}. Private sources of information include offline and online surveys, in-store and online transactions, website and forum interactions, and social networking activity~\cite{experianmarket2013}. Data aggregators  combine the data obtained from these sources and build behavioral profiles of individual users. The data and behavioral profiles are traded on data marketplaces~\cite{bluekai2014market}. Service providers can purchase data and profiles, and use it to enrich their knowledge about their customers, which may help them to improve their services.  

\begin{figure*}
\begin{center}$
\begin{array}{ccc}
\includegraphics[trim= 0 8 0 2,clip,width=2in,height=1.5in]{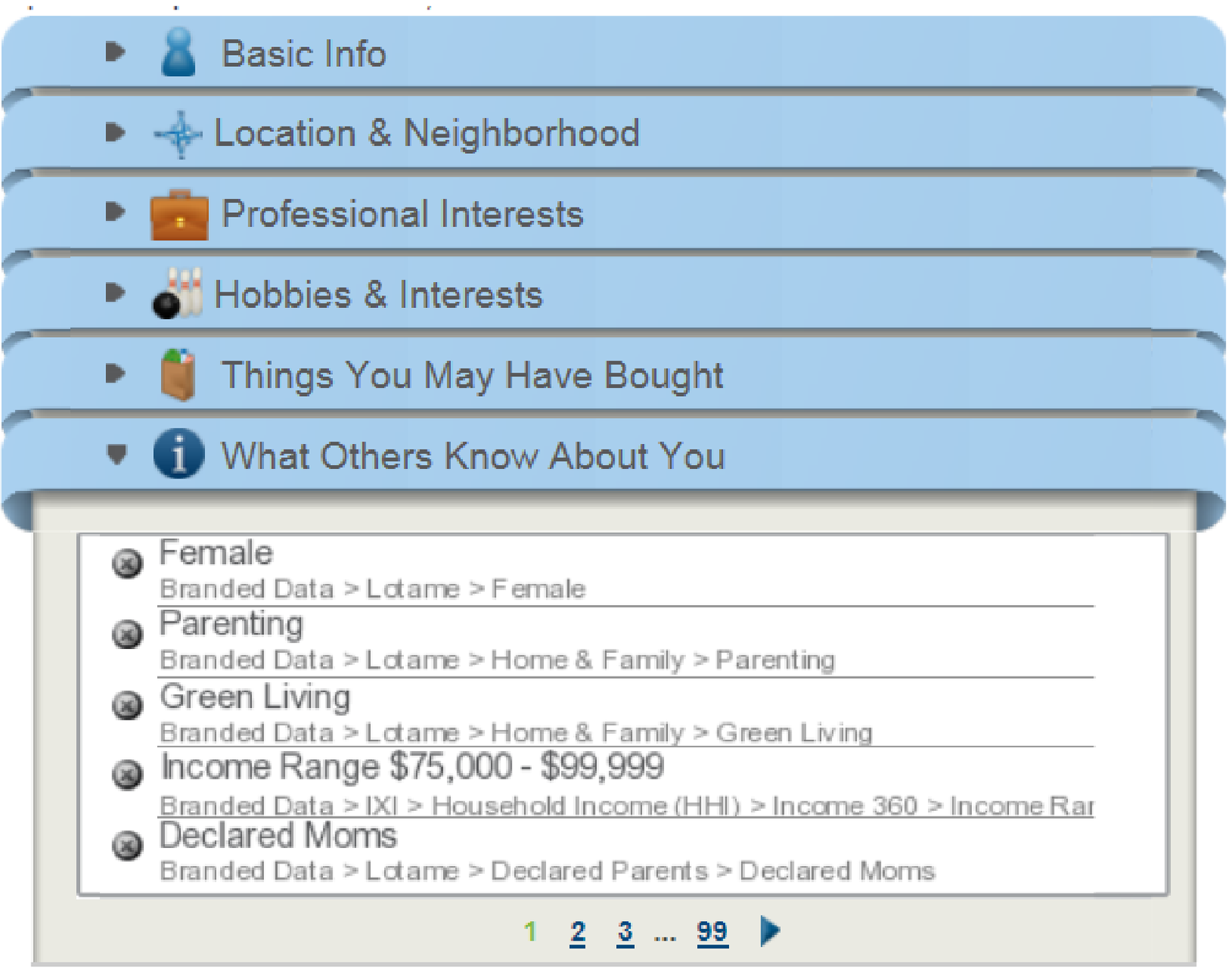} &
\includegraphics[trim= 0 20 10 0,clip,width=2in,height=1.5in]{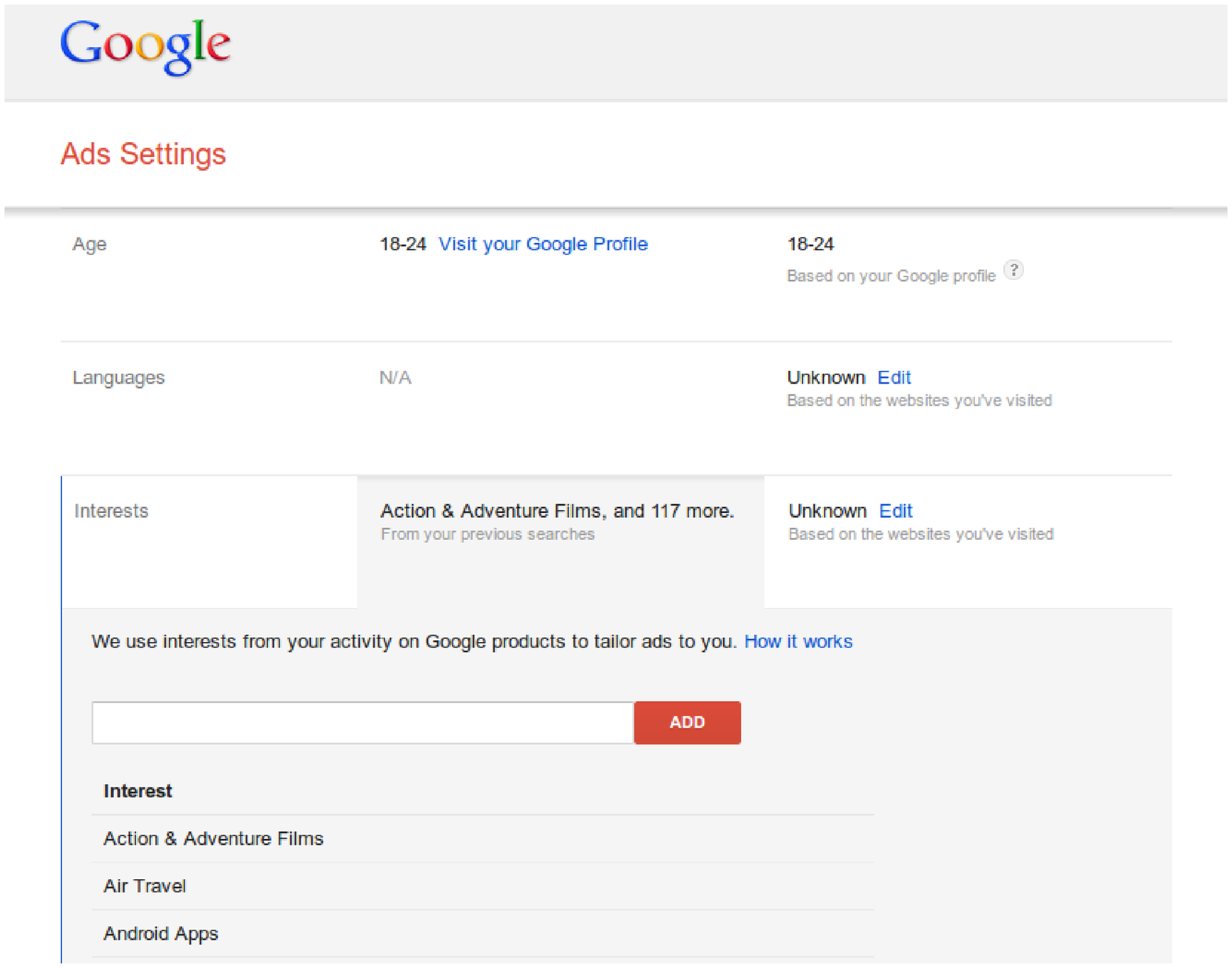} &
\includegraphics[trim= 0 20 10 0,clip,width=2in, height=1.5in]{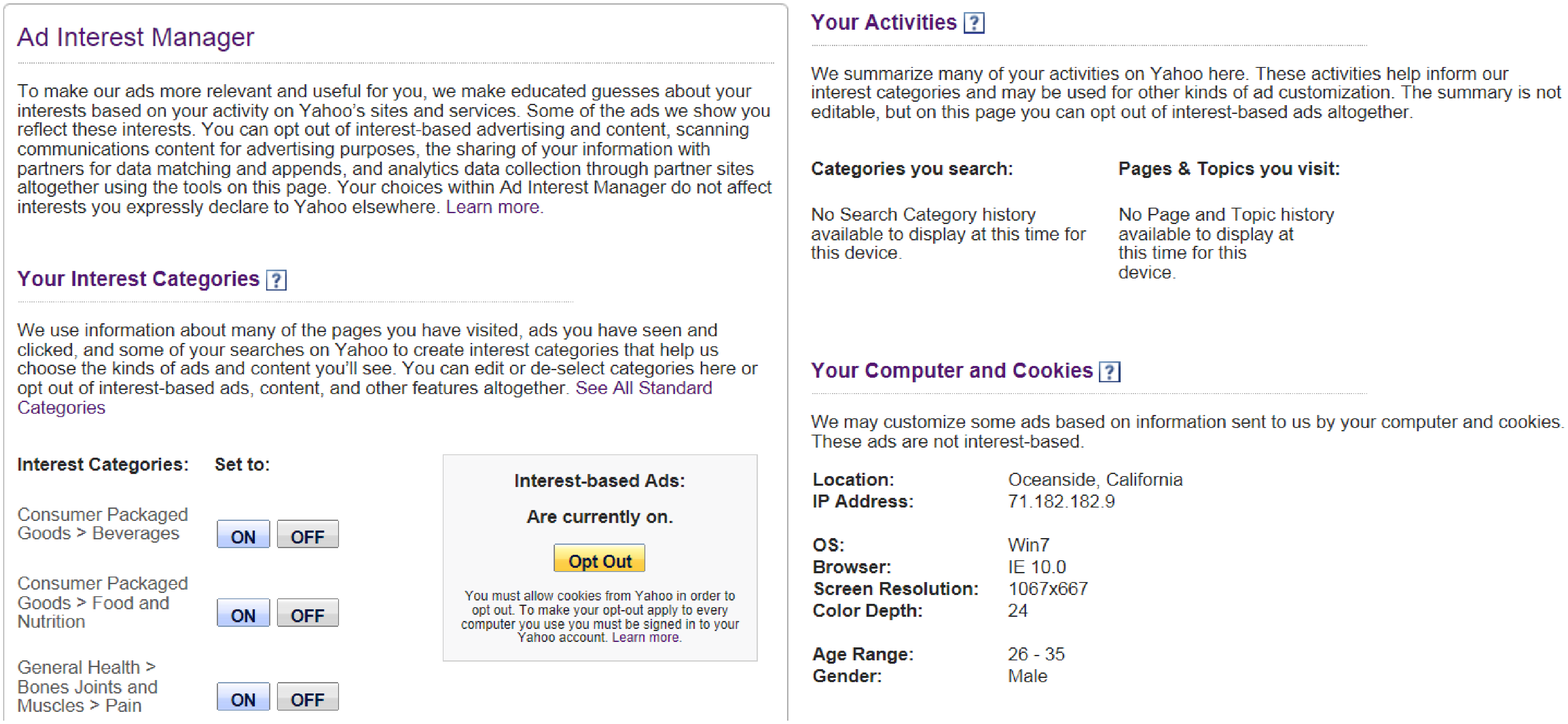} 
\end{array}$
\end{center}
\caption{Sample profiles: BlueKai Registry (left), Google Ad Settings (middle), and Yahoo Ad Interests (right)}
\label{fig:profiles}
\end{figure*}

\paragraph{Accessing Online Behavioral Profiles:}
\label{sec:access}
To increase transparency, some companies allow users to access their behavioral profiles. Companies may choose to show only some of the data that they have collected about the user~\cite{acxiom2014atd}. In addition to looking at their data, users may be able to edit data in their profiles. Companies may use client-side or server-side validation to provide access to user profiles. For example, BlueKai~\cite{bluekai2014registry}, Google~\cite{google2014adsetting} and Yahoo~\cite{yahoo2014adinterest} provide access to profiles based on browser cookies. Companies such as Acxiom~\cite{acxiom2014atd} and Microsoft~\cite{microsoft2014optout} require that users create an account with them and sign-in to access their profiles. To create an account users may have to provide their email address and name. Additionally, companies may request information such as full legal name, full address (street, city, state and Zip code), date of birth and last four digits of social security number to verify the identity of a user~\cite{acxiom2014atd}. In Fig.~\ref{fig:profiles}, we show examples of the three profiles, BlueKai Registry, Google Ad Settings, and Yahoo Ad Interests, used in our study. 

\section{Related Work}
\label{sec:related}
The Federal Trade Commission recently released a comprehensive report on the activities of data aggregators (brokers)~\cite{ftc2014broker}. The report details how data is acquired using various data sources and collection techniques. It discusses the types of data collected, potential uses, and steps taken by aggregators to maintain data accuracy. The report does not investigate contents of actual profiles.   

Several studies have looked into how advertisers use different technical measures such as browser cookies, flash cookies and Javascript to collect different types of data~\cite{hoofnagle2012behavioral, krishnamurthy2009privacy} and track user activities~\cite{mayer2012third, roesnertracking2012}. These studies have largely focused on data collection from online sources and not on data collection from public and offline sources. 

Prior research has studied user understanding, perceptions and concerns regarding targeted advertising, which uses behavioral profiles to personalize ads. Turow et al. surveyed Americans' attitudes towards targeted advertising that used data collected from online websites and offline stores~\cite{turow2009americans}. They used telephone interviews and closed-ended questions to understand attitudes of a representative sample of the US adult population. McDonald and Cranor studied users' understanding of targeted advertising and technical mechanisms such as cookies used for targeted advertising, and user concerns regarding targeted advertising~\cite{mcdonald2010beliefs}. Ur et al. studied user beliefs, attitudes and concerns regarding targeted advertising using semi-structured interviews~\cite{ur2012smart}. Agarwal et al. studied users concerns regarding embarrassing and suggestive ads that may arise out of targeted advertising~\cite{agarwal2013not}. Gomez et al. studied user concerns regarding advertiser data practices by looking at three sources of information: consumer complains to the FTC and other organizations, results from user surveys regarding privacy, and published news articles~\cite{gomez2009knowprivacy}. Kelley et al. studied concerns about location-based advertising and identified different factors that influence users' level of concern~\cite{kelley2011users}. These studies have not investigated privacy concerns regarding actual contents of behavioral profiles, and they have not employed users' own behavioral profiles.

\section{Methodology}
\label{sec:methodology}
In this section, we provide an overview of our study. We explain our choice of behavioral profiles. Lastly, we discuss details of in-person interviews we conducted.

\subsection{Overview of the Study}
We first conducted in-person interviews where we asked our participants to access their own profiles. From the interviews, we gathered and categorized the information observed in the behavioral profiles of our participants. During the interviews, we also elicited participants' concerns and surprises regarding information in their profiles. We discuss details of the in-person interviews below. 

After conducting in-person interviews, we conducted an online survey. We designed the survey to achieve two goals. First, we wanted to confirm whether a more diverse
population of users agreed with the concerns that we had identified from the interviews. Second, we wanted to identify additional user concerns. We defer discussion of the online survey until we have discussed our findings from the in-person interviews. Details of the online survey are in Section~\ref{sec:survey}.

\subsection{Selection of Behavioral Profiles}
We studied behavioral profiles from three companies: BlueKai Registry, Google Ad Settings, and Yahoo Ad Interests (see Fig.~\ref{fig:profiles}). As discussed in Section~\ref{sec:access}, these are cookie-based profiles that do not require users to create accounts or signin on data aggregator websites. We felt participants would find cookie-based profiles easier to access. 

We chose profiles that covered large number of users and contained data from multiple sources. By doing so, we expect our results to be more representative. Data in the BlueKai Registry profiles come from nearly 30 third-party companies that participate in the BlueKai Audience Data Marketplace~\cite{bluekai2014market}. Currently the marketplace is the world's largest third-party data marketplace providing data on 300 million users or approximately 80\% of the US population. Google Ad Settings displays interests and other information inferred from user activities on Google and more than one million partner sites~\cite{google2014adsetting}. Yahoo Ad Interests shows data inferred from Yahoo's sites and services~\cite{yahoo2014adinterest}.

\subsection{In-person Interviews}
We conducted semi-structured in-person interviews with eight participants. We explained to the participants that companies may collect data about them, and may create behavioral profiles. We informed the participants that they might be able to access their profiles. We requested them to look at their profiles (BlueKai, Google and/or Yahoo), and if they felt comfortable, share information in their profile with us. We asked them to voice any concerns, surprises or questions regarding the data in their profiles.

\paragraph{Participant Background:}
Our participant pool included graduate students with engineering and/or science background. Only one participant was aware that he could access profiles created by companies. Six participants had never deleted cookies from their browser, one deleted cookies selectively, and one regularly deleted cookies.  

\paragraph{Data Collection:}
From our eight participants, we collected information on eight profiles including five BlueKai Registry profiles, two Google Ad Settings profiles, and one Yahoo Ad Interests profile. All of the participants tried to access their BlueKai profile. Six were able to access their Bluekai profile, but two were not able to access it as they were using cookie blocking and/or script blocking. Of the six participants who accessed their Bluekai profile, five shared profile information with us. Three of our participants looked at an additional profile; two looked at Google profile and one looked at Yahoo profile, and they shared profile information with us.   

\paragraph{Data Collection Challenges:}
Seven of our participants showed us contents from their profiles. Six of them pointed out specific items from their profiles that concerned or surprised them, but did not share the entire profile with us. The primary reason for this was the time it took to share the entire profile. Due to the way the profiles were displayed, it was not possible to download entire contents of a profile into a spreadsheet or XML document. To share information, a participant had to take screen shots of each page in the profile. When there were multiple pages, for example, 99 pages in the sample BlueKai profile in Fig.~\ref{fig:profiles}, participant had to click through the pages. Further, individual entries in a page in a BlueKai profile were images and not text, and, hence, it was not possible to copy and paste entries into a text document. Since we studied cookie-based profiles, participants were accessing profiles from their work or personal computers, and most did not want us to access it in their absence. One participant, who had an extensive profile, provided us access to his computer. It took us over an hour to copy the entire profile. This is one of the usability issues in accessing profiles.        

\section{Contents of Behavioral Profiles}
\label{sec:contents}
We discuss the types of data found in behavioral profiles of our interview participants. Further, based on the information possessed by companies that feed data into BlueKai profiles, we discuss what other types of data may be found in behavioral profiles.  

\subsection{Analysis of Profiles}
We analyzed the eight profiles from our interview participants for different types of data. We had profiles from three different companies, Bluekai, Google and Yahoo. We computed size or the total number of data items in each profile. The two Google profiles had $\simeq$120 items, the Yahoo profile had $\simeq$25 items, one Bluekai profile had $\simeq$10 items, two BlueKai profiles had $\simeq$30 items, and two BlueKai profiles had $\simeq$570 items. Based on the number of items, we can say that we had four small-sized profiles, two medium-sized profiles and two large-sized profiles. 

We organized the data from these profiles into seven categories: demographic, geographic, technical, predictive, psychographic, behavior and life event. We based our categories on the categories commonly used in data marketplaces and privacy policies to describe different types of data. We tried to choose distinct, non-overlapping categories, so that each data item would fall into only one category. However, for some data items, it was difficult to choose a category. For example, it was difficult to decide whether ``Credit Card Holder'' belonged to individual demographic or behavioral data category.   

A challenge during analysis was to comprehend profile items. For example, the meaning of ``Demographic $>$ High Confidence'' and ``Credit Card Interest Score'' was not clear. Does ``High Confidence'' imply that the user has high confidence or that the company has high confidence in the accuracy of demographic data? Does ``Interest Score'' mean how much interest a user is paying or how much he is interested in getting a new card? We were able to resolve some of the ambiguities by reading several documents published by data companies. For example, we resolved ``High Confidence'' as implying data accuracy, but were unable to disambiguate the meaning of ``Interest Score.'' Our participants also had difficulty with comprehension. We consider this as another usability issue in accessing profiles. 

Geographic category was present in Yahoo and BlueKai profiles. Only Yahoo profile contained technical category. All three profiles showed individual demographic data regarding gender and age. However, BlueKai profile contained additional individual demographic data including marital status, education and occupation. It also contained demographic data related to user's household and work. The remaining categories appeared only in the BlueKai profiles.

Note that if a profile from a company does not show a certain category, it does not imply that the company does not have such information; a company may choose not to show some of the categories. Yahoo, for example, states on its Ad Interests Manager page ``In addition to the information shown here, Yahoo! may use ... information provided by partners to help customize some of the ads...~\cite{yahoo2014adinterest}.'' Further, ``Yahoo! may combine information, including personally identifiable information, that we have about you with information we obtain from our trusted partners,'' and BlueKai is one of its trusted partners~\cite{yahoo2014append}. In terms of improving transparency by allowing users to see the data in their profiles, BlueKai profiles are better than Google and Yahoo profiles because they provide more detailed information.

\subsection{Profile Contents}
We describe the seven categories of data types we found in actual behavioral profiles. We provide a summary with examples in Table~\ref{tab:data_summary}. For demographic, geographic, technical and life event categories, we describe all the data types we found. However, for psychographic, behavioral and predictive categories, the number of data types that we found are many, and, hence, we discuss representative examples. Further, for each category, we contrast what we found with the data that we may find if we examine more profiles. 

\subsubsection{Demographic data}
Demographic data contains individual, household and firmographic subcategories. Companies associate individual's full name, full postal address, mobile number, email address and email activity date with both demographic and other categories discussed below~\cite{v122014linkage}.

\paragraph{Individual demographic:} We found gender, age (e.g. 20-24 years), marital status, education level (e.g. Some College), occupation (e.g. IT Professional), voter indicator, parent (e.g. Declared Mom), home ownership (e.g. Home owner or Renter) and languages. We found age, but companies also have date of birth~\cite{bluekaibook2013}. In addition to voting, they have party affiliation (e.g. Democrat) and political donor (e.g. Contribute conservative) data~\cite{v122014select,bluekaibook2013}. Other information include religious affiliation (e.g. Hindu), race/ethnicity (e.g. Arabs), family position (e.g. Female head of household) and summarized credit statistics including wealth rating (e.g. Decile), credit rating (e.g. High) and net worth~\cite{v122014select,experianmarket2013,bluekaibook2013}. 
\paragraph{Household demographic:} It includes details of an individual's household. We found household income (e.g. \$20K-\$30K), household size (e.g. 1), number of adults (e.g. 1), children in household (e.g. No), home type (e.g. Multifamily Dwelling), median home value (e.g. \$0-\$100K), length of residence (e.g. Less than 3 years), discretionary spending (e.g. \$40K-\$50K) and auto (e.g. Less than \$20K). In individual demographic, we did not find individual income, but when household size or number of adults is one, then household income implies an individual's income. 

For household demographic, companies have rich set of additional attributes. In addition to knowing presence of children in a household, they know number of children, their gender and age, which can be a range (e.g. 0-3 years, 4-7 years) or month, day and year of birth~\cite{experianmarket2013,bluekaibook2013,v122014select}. They have indicators for the types of persons in a household, for example, presence of smoker, veteran in household, elderly parent in household~\cite{experianmarket2013}. Further, they have data about mortgage and refinance (amount, term, loan type, rate type)~\cite{v122014select}.

\paragraph{Firmographic:} It generally includes details about an individual's profession and affiliated organizations. We found type of industry (e.g. College and University), number of employees (e.g. 1-20 employees), and characteristics of the profession (e.g. High Net Worth) and position (e.g. Technical Business Decision Maker). Additionally, companies have data about sales revenue, years of establishment (e.g. $<$2 years), domain expertise and seniority~\cite{bluekaibook2013}.

\subsubsection{Geographic data}
Geographic data includes location and neighborhood of a user. For example, we found ``US $>$ Pennsylvania $>$ Pittsburgh,'' ``US$>$Massachusetts$>$Boston-Cambridge-Quincy'' and ``Oceanside, California'' for a participant that currently lived in Pittsburgh and Boston, and had lived in Oceanside about five years ago. The smallest granularity we found was at the city/county level. However, companies have geographical data at the level of full postal address, Zip code +4 (block level) and Zip code~\cite{bluekaibook2013,v12postal2014}. For example, one company from BlueKai Marketplace claims to have 208 million postal addresses~\cite{v12postal2014}, and 72 million postal addresses linked to email addresses~\cite{bluekaibook2013}. Each postal record is linked to a consumer's demographic, interests and behavioral data.  

\subsubsection{Technical data}
Technical data generally includes information related to users' computers and devices used to access the Internet. We found IP address (e.g. 71.182.182.9), operating system (e.g. Windows 7), browser (e.g. IE 10), color depth and screen resolution. Interestingly, companies may use IP address to identify an anonymous consumer visiting a website in real-time. For example, they can map an IP address to a consumer's full name, full postal address, mobile number, purchases, interests and $\simeq$260 more attributes~\cite{v122014linkage}. They also use IP address to infer location, for example, Yahoo states, ``We use the IP address to infer your location ...''

We did not find technical data regarding browser cookies and online activities and interactions, for example, search history, websites visited, articles read, comments, ratings and uploaded files. However, companies collect such information to derive psychographic, behavioral and predictive data. They use browser cookies to identify a website visitor's gender, presence of children (Yes or No), age (e.g. 20-29) and household income (e.g. $75,000-$99,999)~\cite{bk2014demographics}. 

\subsubsection{Predictive data}
Companies generally employ proprietary models and algorithms that combine data from multiple public, proprietary and self-reported sources, both online and offline, to make predictions about users. Predictions can be made about behavior, attitude, interest etc. For our predictive data category, we consider data that indicates user's intent to purchase, usually in the near future. We discuss other types of predictions as part of other categories discussed below.  

We found examples that predicted purchases related to credit card, personal health, higher education, computers, cell phones, auto insurance, flying, hotels etc. For example, ``Credit Card App Intent Score -- 10-11\%'' indicates intent to apply for a credit card. ``Personal Health -- Values 70-90\%'' indicates future purchase propensity regarding personal health products; ``In-Market -- Cell-Phones and Plans'' and ``In-Market -- US Domestic Flyers'' indicate that the user is currently shopping for cellphone plans and flights; ``Auto insurance online buyer -- High Propensity'' and ``Online Higher Education Enrollee -- High Propensity'' indicate users looking to buy insurance or enroll in courses. Companies have in-market data for many other areas including real estate, apartments and automotive purchases~\cite{bluekaibook2013}.

\begin{table}
\renewcommand{\arraystretch}{1.1}
\caption{Examples of Data Types Found in User Profiles}
\label{tab:data_summary}
\centering
\begin{tabulary}{0.45\textwidth}{l L}
\hline
Category  & Examples \\
\hline
Demographic             & \\
	\multicolumn{1}{l}{~~~Individual}		& Female \\
						 						& Single \\
						 						& 20-24 years \\
		  				 				  & Some College \\
						 						& IT Professional \\
						 						& Voter\\
	\multicolumn{1}{l}{~~~Household}  & Income Range -- \$20K-\$30K \\
						 						& Household Size -- 1 \\
						  						& Children in Residence - No \\
						  						& Home Value -- \$0-\$100K\\
						 						& Length of Residence -- Less than 3 years \\
						  						& Auto -- less than \$20K \\
	\multicolumn{1}{l}{~~~Firmographic} & Business Data $>$ Micro (1-20 employees) \\
						               & Business Data $>$ Software \\
\hline
Geographic   & US $>$ Pennsylvania $>$ Pittsburgh \\
					 													& Oceanside, California \\
\hline
Technical   													& IP address -- 71.182.182.9 \\
					 													& OS -- Win7 \\
					 													& Browser -- IE10 \\
					 													& Screen resolution -- 1067X667 \\
\hline
Predictive     					& Credit Card Interest Score -- 16-17\% \\
													& Credit Card App Intent Score -- 10-11\% \\
							           & Auto insurance online buyer -- High Propensity \\
							           & In-Market -- Cell-Phones and Plans \\
\hline
Psychographic 				 &  \\
 \multicolumn{1}{l}{~~~Interests}   	& Health $>$ Bones, Joints, Muscles $>$ Pain \\
                         & Interest in Religion -- Value Tiers 1-3 \\
							           & Sweepstakes -- Value Tiers 1-3 \\
													& Weight Conscious Code - Value Tiers 1-3 \\
						   					& Travel Destinations $>$ New York \\ 
\multicolumn{1}{l}{~~~Attitudes}	 & Buy American -- Not Likely \\
							 					& Show me the Money -- Most Likely \\
\hline
Behavior			 & \\
		\multicolumn{1}{l}{~~~Activities} & OTC Medicine $>$ Pain Reliever \\
							 					 & Gastrointestinal -- Tablets \\
							 					 & Offline CPG Purchasers $>$ Brand $>$ Hebrew National \\
						   					 & Charmin Ultra Soft \\
													 & Past purchase $>$ ISP $>$ Internet $>$ Verizon \\
		\multicolumn{1}{l}{~~~Lifestyle}	& Green Living \\
													& Owns a Regular Amex Card \\
													& Eco Friendly Vehicle Owner \\
							 					& Premium Channel Viewer \\
\hline
Life Event               & Empty Nesters \\ 
\hline
\end{tabulary}
\end{table}

\subsubsection{Psychographic data}
Psychographic data generally includes interests and attitudes of a user. We found interests related to health (e.g. Bones, Joints, Muscles $>$ Pain, Weight Conscious Code -- Value Tiers 1-3), religion (e.g. Interest in Religion Code -- Value Tiers 1-3, Christian Music Code -- Value Tiers 1-3), travel (e.g. Destinations $>$ New York, Vacation Packages), automotive (e.g. Coupe), sweepstakes, news (e.g. News and Politics $>$ Government) etc. Companies possess additional data including gambling, lottery, alcohol and tobacco~\cite{experianmarket2013}.

Profiles can include data on attitudes and values of users. Companies can use that information to trigger desired response from users. Some of the attitudes we found are as follows. ``Buy American -- Most Likely,'' which may indicate relatively high importance of pride in decision making. ``Work Hard, Play Hard -- Not Likely,'' which may indicate users' desire to be at the forefront of both their career and outside relative to their peers. ``Stop and Smell the Roses -- Most Likely,'' which may indicate a belief in altruism.

\begin{figure}
\centering
\includegraphics[trim=0 0 0 0,clip,width=2.5in]{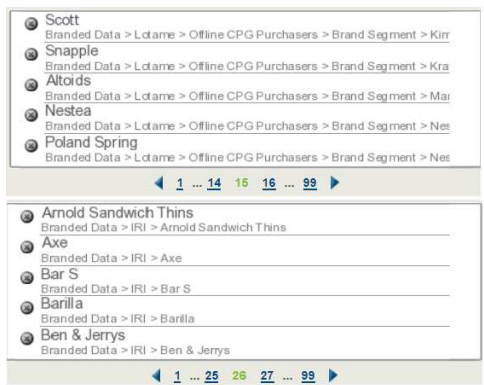}
\caption{Listing of Consumer Packaged Goods in a profile}
\label{fig:cpg_combo}
\end{figure}

\subsubsection{Behavioral data} 
Behavior data contains data related to users' lifestyle, activities and personality. For example, the entry, ``Green Living,'' found in one of the profiles indicates that the user exhibits an environmentally friendly lifestyle. Companies can further differentiate between users who act and those who only think (e.g. Behavioral Greens vs. Think Green), and between undecided and those who are against (e.g. Potential Green vs. True Brown)~\cite{experianmarket2013}. The profile containing ``Green Living'' also contained ``Eco Friendly Vehicle Owner.'' Other lifestyle aspects we found include credit (e.g. Owns a Regular Amex Card), finance (e.g. Owns Mutual Funds), shopping (e.g. Discount Shopper) and travel (e.g. Theme Park Visitor).  

In activities, we consider past purchases, both offline, such as stores and pharmacies, and online. One participant had over-the-counter medications (e.g. OTC Medicine $>$ Pain Reliever, OTC Medicine $>$ Cough and Cold) purchased at a local pharmacy listed in her profile. In addition to OTC, companies have information about medications (e.g. oral contraceptive, Lipitor, Insulin) purchased by users and any ailments they may have (e.g. Alzheimer's, clinical depression, Diabetes-2)~\cite{experianmarket2013}.

Another participant had a list of $\simeq$300 past consumer packaged goods (CPG) purchases in his profile. We list part of his profile in Fig.~\ref{fig:cpg_combo}. CPG entries include brands (e.g. Hebrew National, H$\ddot{a}$agen-Dazs) and items (e.g. Charmin Ultra Soft, Gastrointestinal -- Tablets, General Mills $>$ Fiber One). Companies have other data such as purchase of alcohol and tobacco~\cite{experianmarket2013}. Lastly, companies have built models to predict an individual's personality type (e.g. introvert, leader). They have assigned personality types, by name and postal address, to 85\% of the US adult population~\cite{v122014pyco,bluekaibook2013}.

\subsubsection{Life event data}
Life event data indicates certain events in a user's life that may lead to changes in behavior and/or create specific needs. We found ``Empty Nester,'' which may indicate that the user's children have left for college. Other life events that companies focus on include new movers and new parents~\cite{bluekaibook2013,experianmarket2013}.   

\section{User Concerns}
\label{sec:concern}
As part of the interviews, we asked participants (n=8) about their concerns and surprises regarding the data in their profiles. Below, we discuss their concerns and surprises. Note that participants viewed their own profiles, which varied among participants. 

\paragraph{Collection of sensitive data:}
Participants expressed surprise and/or concern about credit and health information. One participant was surprised by credit information ``Credit Card App Intent Score -- 10-11\%'' and ``Credit Card Interest Score -- 16-17\%.'' He was concerned because he did not understand the meaning or implication of the credit information in his profile. A participant who found a over-the-counter medication ``OTC Medicine $>$ Pain Reliever'' said that it scared her. She had recently purchased pain medications from pharmacy for an injury. Another participant who had ``General health $>$ bones, joints, muscles $>$ pain'' considered the data confidential and did not want it to be in his profile. As result of an injury suffered during an accident, the participant was in pain for a prolonged time. He had not shared the details with other people. In his opinion, extracting this information from a few online searches and reflecting it in his profile was akin to sharing the information with others. 

\paragraph{Combining data and extent of collection:}
One participant who had an extensive profile with $\simeq$570 items was surprised and concerned by the amount of data gathered. The participant's profile contained demographic -- e.g. age, gender, household income -- location, past purchases including a comprehensive list of $\simeq$300 offline consumer goods purchases etc. The participant was surprised about how all the information was obtained without his knowledge or consent. Further, he was concerned to see his data from multiple sources being combined. He explained that it is okay for individual companies to have data about his business with these companies, for example, cellphone company knowing about cell phone plans, or pharmacy knowing about consumer goods purchases. However, a third party combining data from multiple sources and building profiles was not okay to him. The participant mentioned that it was not clear how all this would affect him.  

\paragraph{Granularity of data:}
For some data types, the concern was regarding the granularity or level of detail. One participant was okay with broad interest categories, but not with specific categories. For example, he was not concerned to see ``Web Services'' listed under interests. However, he would be concerned if a specific instance such as ``Pirate Bay'' was listed. Another participant was concerned about granularity of retention period. He pointed out that a health condition listed in his profile was more than five years old. The participant had forgotten about it, but the information was still present in his profile. The participant's concern is similar to the ``right to be forgotten'' argument~\cite{rosen2012right}.      

\paragraph{Data use:}
Participants were concerned about how the data in their profiles may be used. One of the participants, who had credit scores listed in his profile, was concerned about its implications. One more participant expressed similar sentiment when he said it was not clear how the extensive collection and combining of data would affect him. Both the participants were indirectly, if not directly, thinking about the purposes for which the data may be used. Another participant was more direct: he felt that data can be used to infer actions performed by the user. He was concerned that by combining different interests, for example, Pirate Bay and Movies, one could conclude that he had downloaded movies illegally.  

\paragraph{Accuracy of data:}
All profiles had errors to varying degrees, and errors were found among all data types. In general, participants were not concerned when the data was incorrect. A participant even stated that he was happy that there were so many errors. Participants, however, became concerned when the data in the profile was correct. For example, one participant initially found many entries regarding credit and income, but was not concerned. This was because the entries consistently, but erroneously, stated that the participant was affluent with 350000+ income, had top 1\% credit and owned American Express card. However, after seeing an OTC medication entry that was correct, the participant said, ``Now I am scared." Later, this participant hypothesized that companies added incorrect data to profiles so users would not worry too much. A participant expressed concern when only two out of twelve entries regarding professional interests were correct. One reason that contributed to user concern was the level of detail or specificity of the correct entries. Only one participant pointed out that he would be concerned about incorrect data if it was used to make adverse decisions about him. This is interesting as it highlights the importance of accuracy in behavioral profiles as perceived by users.

\paragraph{Editing profile data:}
In general, participants did not try to correct erroneous data in their profiles. Two participants said that correcting the data would enable companies to track them further. A second reason was that the implications of editing data was not clear. One of the participants asked ``What does edit mean? Is the data deleted from all sources?'' However, we hypothesize that users may want to correct the data in their profiles if erroneous data may lead to decisions that adversely impact them. For example, a user interested in loans may want to correct mortgage amount, credit card score or median bankruptcy score if she believes that a loan company may use that data.  

\section{Online Survey}
\label{sec:survey}
We conducted an online survey (n=100) to validate the identified concerns with a larger and more diverse population. This survey had two purposes. First, we wanted to confirm whether a more diverse population of users agreed with the concerns that we had identified from the interviews. Second, we wanted to identify potential additional user concerns and data types that may not have been observed in the interviews. We recruited survey participants from Amazon Mechanical Turk crowd-sourcing platform~\cite{mturk2014}. We provide the survey questionnaire in Appendix~\ref{sec:appendix}. 

\subsection{Survey Design}
To understand participant demographic, we asked them their age, gender, primary occupation and education level. To understand their technical background, we asked them whether they had a college degree or work experience in computer science, software development, web development or similar computer-related fields. We also asked them how much they liked personalization of ads on websites. We gathered information on demographic, background and liking for personalization as they may affect participant concerns. We also used demographic data to analyze diversity of our participant population.

\begin{figure}
\centering
\includegraphics[trim=0 0 0 0,clip,width=2.5in]{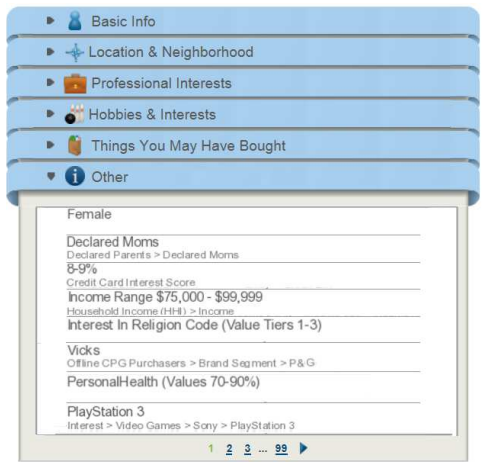}
\caption{Sample profile used in online survey}
\label{fig:sample_profile}
\end{figure}

We used a sample profile shown in Fig.~\ref{fig:sample_profile} to understand whether the survey participants agreed with the concerns that we identified from the interviews. We used the sample profile to understand their concerns regarding collection of sensitive data, amount of data, combining data from multiple sources, level of detail and data use. We felt that survey participants could not provide meaningful answers regarding concerns of accuracy of information and editing profile data based only on a sample profile. Hence, we did not ask them about those concerns.

We created the sample profile using data from profiles of the interview participants. To understand concerns about sensitive data collection, we added items related to credit (Credit Card Interest Score 8-9\%) and health (Personal Health -- Values 70-90\%) both of which our interview participants had found sensitive. We also added entries related to religion (Interest in Religion Code -- Value Tiers 1-3), individual demographic (Female and Declared Mom) and household demographic (Income Range \$75K-\$99K). To address the concern on amount of data, we ensured that the profile had data items from several categories: demographic, psychographic, behavior and predictive. Geographic category was represented by the ``Location and Neighborhood'' tab. To show data being combined from multiple sources, we added an offline CPG purchase (Offline CPG Purchasers $>$ Vicks). To cover concern about level of details, we picked items that were very specific ``Interest $>$ Video Games $>$ Sony $>$ PlayStation 3.'' Further, the predictive values such as ``Values Tiers 1-3'' also increased the specificity of items. Lastly, we felt that it would be more realistic to show the data items as they appeared in actual profiles; a user looking at her actual profile would not have additional explanations or links to documents that could clarify her ambiguities.        

Before showing the sample profile, we explained to the participants that advertisers collected data about them in order to personalize ads. Further, advertisers may create profiles about them using the collected data. We then showed them a sample profile (Fig.~\ref{fig:sample_profile}). To check whether participants were paying attention, we asked them to select, from a list of six items, at least two items present in the sample profile. We then asked the participants to rate, on a 5-point Likert scale of ``Strongly disagree'' to ``Strongly agree,'' how much they agreed or disagreed with the following list of concerns. We randomized the order in which the concern statements were displayed. 
\begin{enumerate}
	\item I am concerned because I believe that the profile contains sensitive data
	\item I am concerned by the amount of data in the profile
	\item I am concerned because my data from multiple sources (e.g. online activities, in-store, other companies) is being combined
	\item I am concerned by the level of detail (e.g. specific information, not just broad categories) in the profile
	\item I am concerned about how my data may be used
\end{enumerate}

After the participants rated the concerns, we asked them, using an open-ended question, whether they had any other concerns regarding the sample profile. We also asked them, using a 5-point Likert scale, if their liking for personalization had decreased after seeing the types of data collected for personalization. We were interested in knowing if awareness of behavioral profiles can change participants' opinions.

Since we could not address, with a sample profile, concerns regarding accuracy of information and editing profile data, we gave participants the option of looking at their own profiles. We made this step optional, to know whether participants were really interested in looking at their own profiles. We stated that their payment and bonus were not affected if they chose not to look at their profiles. For participants who chose to look at their own profiles, we provided instructions to access BlueKai, Google and Yahoo profiles. We then gave these participants an option to describe their reactions. This also helped us identify any additional concerns or data types. Lastly, we asked all participants if they had any further comments.      


\subsection{Participant Background}
We recruited participants (n=100) from Amazon Mechanical Turk crowd-sourcing platform~\cite{mturk2014}. Our participants were at least 18 years of age and located in the United States. We used the Mechanical Turk location feature to ensure that users were from the United States. We collected informed consent from our participants. We offered a payment of \$0.5 for completing the survey and a \$0.3 bonus for following the survey instructions correctly. We implemented our survey on the Survey Gizmo platform, and redirected participants from Mechanical Turk to Survey Gizmo.

The average age of the participants was 27.74 years ($SD = 7.57$) and median was 26 years. The male to female ratio was four to one. Thirty seven participants had completed a four year bachelors degree or higher. Twenty five participants had a college degree or work experience in computer science, software development, web development or similar computer-related fields. Twenty five participants were students, and the rest had diverse occupations including administration, art, business, education, engineer, law enforcement, service, skilled labor and homemaker. Our survey participant pool was more diverse than our interview participant pool especially in education level, occupation and technical background. Thirty six participants agreed (8 strongly agree, 28 agree) that they liked personalization of ads, and 39 disagreed (12 strongly disagree, 27 disagree). Hence, the pool was balanced in its opinion of personalization. 

\subsection{Survey Results}
Out of the 100 participants who completed the survey, 97 correctly answered the attention check question, that is, they selected at least two items present in the sample profile. Two participants selected one item not present in the profile, and one selected two items not present in the profile. 
 
\begin{figure}
\centering
\includegraphics[width=1.5in, angle=270]{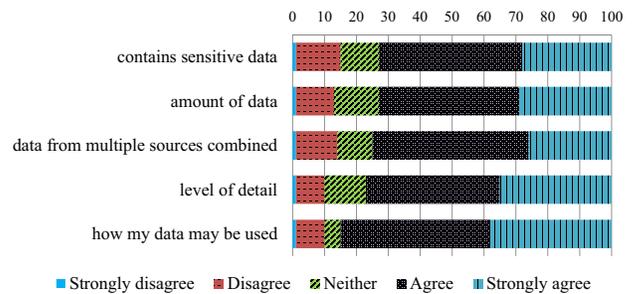}
\caption{Percentage (x-axis) of survey participants (n=100) who agreed with indicated concerns (y-axis)}
\label{fig:user_concerns}
\end{figure}

In Fig.~\ref{fig:user_concerns}, we show how survey participants (n=100) rated concerns regarding collection of sensitive data, amount of data, combining data from multiple sources, level of detail and data use. For each of the five concerns, at least 70\% of the participants either agreed or strongly agreed that they were concerned. Participants were most concerned about how their data may be used (85\%), followed by level of detail (77\%), aggregation (75\%), amount of data (73\%) and collection of sensitive data (73\%). Using a MANOVA, we found that the differences among user concerns were significant ($F[4, 96] = 3.9, p < .05$). At least 70\% agreement on each concern assures us that a more diverse population agrees with concerns that we identified from our interviews.  

We analyzed participant comments for additional concerns. Majority of the participants did not express new concerns. Seven participants were concerned about the security of their data; they worried that their data could be abused by hackers, criminals and identity thieves. Four participants expressed concerns that their data could be shared or sold to third parties, and accessed by the government. These are important and should be explored further. 

Fifty participants agreed (17 strongly agree, 33 agree) that their liking for personalization had decreased after seeing the types of data collected for personalization, and 23 disagreed (2 strongly disagree, 21 disagree). Interestingly, 18 of those 50 participants were participants who liked personalization of ads.
 
Seventy one participants (71\%) chose to look at their own profiles even when it was optional. This indicates that people are interested in learning about their behavioral profiles. This may also indicate that many people are unaware of profile access mechanisms provided by companies. This is similar to our interview pool where only one out of eight participants was aware of profile access mechanisms. 

Out of 71 participants, 51 (72\%) chose to report their reactions. We analyzed their comments for concerns regarding accuracy and editing profile data. Nine participants (17\%) reported empty profiles. Twenty three participants (45\%) reported inaccuracies, and only three participants (6\%)) reported that they found accurate profiles. Participants reactions to inaccuracies included ``blatantly incorrect,'' ``80\% inaccurate,'' ``somewhat dated'' and ``hilariously overestimated.'' Recall that all our interview participants had also found varying levels of inaccuracies in their profiles. Most of the participants who reported inaccuracies and empty profiles explained that they felt relieved and less concerned about data collection. Only two participants (4\%) felt that inaccuracies in their profiles could adversely affect them. Three participants mentioned about editing data. One of them corrected errors, and two of them deleted correct entries. Reactions of survey participants regarding inaccuracies in profiles and editing profile data appear similar to those of interview participants. 

During analysis of participant reactions, we did not find any new data types. Lastly, we looked for comments that signaled difficulty in comprehending profile information. One participant explicitly reported not being able to understand parts of his BlueKai profile. Two participants thought ``High/Medium Confidence'' was referring to their personality. Some of our interview participants had similar difficulties. Overall, our survey results confirm the results from our interviews.   


\section{Limitations}
We studied profile contents from relatively small number of profiles (n=8). We looked at behavioral profiles from three data aggregators, and all of them were cookie-based profiles. If we study larger number of profiles, profiles from other companies, or server-based profiles, we may find other types of data. We conducted in-person interviews with graduate students (n=8) with science and engineering background. For our online survey, we recruited participants (n=100) from Amazon Mechanical Turk, and they may have more technical knowledge than an average person. Further, our online survey results may contain self-selection bias. By recruiting participants from a more diverse pool, we may identify new concerns and surprises. Lastly, we can improve estimation of profile accuracy by asking participants to verify information on all entries in their profiles.

\section{Discussion and Conclusions}
\label{sec:discussion}
Below we discuss insights from our study. We conclude with a discussion of implications for public policy management and directions for future research.
 
\paragraph{User concerns are justified:}
Our study shows that participants have several concerns about behavioral profiles including extent of collection, collection of sensitive and confidential data, and level of detail. Our interview participants considered health and credit data sensitive. Profiles contained other data such as religion and income types, which a more diverse audience may find sensitive. Further, our analysis of data aggregator documents shows that they have much more intrusive data including fully identifying data such as first and last names, and complete postal addresses. This can further exacerbate user concerns. 

\paragraph{Clarifying data usage is essential:}
The biggest user concern was how their data may be used. Use of profile data is not clear. Given the variety of data present in the profiles, its uses seem limitless. Data could be used for personalization, development of better products, or fraud detection. It could also be used for hiring decisions, discreet background checks or proselytizing. For a user, the impact of using her data for the former could be quite different from that of the latter. 

An important underlying issue is what inferences are permissible. The richness of profile data allows one to draw all kinds of inferences about a user. If a user liked race cars on Facebook, is he likely to speed? If a user brought OTC pain medications frequently, is she addicted to pain medications? Is a user who purchases LeanCuisine brand more healthy than a user who purchases H$\ddot{a}$agen-Dazs brand? Is a user who regularly buys HebrewNational brand Jewish? 


\paragraph{Claims of anonymity of profiles are misleading:}
Companies overlay anonymous data such as financial records with identifying information obtained from public, online and offline sources. This action of combining information from multiple sources not only creates a rich, 360-degree view of all aspects of life, but also associates it with a specific individual, her name, address and other personal information. Statements that imply that profile data are anonymous or pseudonymous, for example ``Consumers can also control their anonymous profile~\cite{bluekai2014registry},'' are misleading. 


\paragraph{Accuracy of profiles is poor:}
Our study shows that a large number of behavioral profiles contain inaccuracies. All interview participants (8/8) and 45\% (23/51) of survey participants, who provided feedback about their profiles, reported errors. This violates an important fair information practice principle: the data quality principle. Although companies seem to be verifying the accuracy of the data that they obtain~\cite{bluekaibook2013}, it is not clear how effective their processes are. Since data is being combined from multiple companies, a few companies taking steps to ensure correctness may not be sufficient. 

Some companies claim that their sources are accurate as they are ``self-reported'' by users and not modeled or predicted. The correctness of these self-reported sources are questionable. Users may be taking surveys or registering without being aware of implications in a different context. In fact, research has shown that people deliberately provide fake data as a way of protecting their privacy online~\cite{pew2013anonymity}. There are many other ways in which errors may be introduced: sharing a store loyalty card with another shopper who forgot her card, browsing from a friend's account, or purchasing items for your employer. 

It is also important to consider the accuracy of predictive data. It is debatable how accurate the results are when a company predicts religious affiliation, country of origin, ethnicity and languages spoken, based on an individual's name~\cite{experian2013ethnic}. Further, desired level of accuracy would depend on the type of data (likelihood of buying toilet paper vs. median bankruptcy score) and its potential uses (advertising vs. hiring).

Interestingly, users were generally not concerned to see inaccuracies. Many felt relieved and did not want to correct the errors. Users appeared to be thinking mainly about companies tracking them, and having incorrect information about users seemed to defeat that purpose. However, users also worried over how their data may be used. Decisions based on erroneous data, for example, fraud detection based on incorrect purchases or job screening using incorrect personality type, may adversely impact users. Hence, we hypothesize that users will start caring about inaccuracies as they become more aware of its implications. 

\paragraph{Effect of editing/deleting profile data is unclear:}
Some study participants deleted data from their profiles to ensure that companies no longer have data about them. Are edit mechanisms meeting this expectation? There are several questions about the effect of editing or deleting profile data. Do all companies that possess a user's data honor a user's request? For example, BlueKai profile shows data that its affiliates may have about the user. Does deleting data from a BlueKai profile guarantee that the data is deleted from its affiliates databases? When a user corrects an erroneous entry in a profile, is that information propagated to companies that acquired the profile data? We need to clarify the implications of edit and delete. Otherwise, they only provide a false sense of comfort to users.

\paragraph{Transparency provided by access is insufficient:}
From our study, we believe that providing access to user behavioral profiles is a step in the right direction; it improves transparency of data practices. However, the information provided via these access mechanisms is incomplete and insufficient. First, our study shows a large gap between the types of data companies show in user profiles and data that they actually possess about users. For example, profiles show age, but companies also have date of birth; profiles show city, but companies also have Zip, Zip+4 and postal addresses; and companies state profiles are anonymous, but they have full names. Second, some companies that provide access are more transparent than others, for example, BlueKai vs. Yahoo or Google. Lastly, profiles show information about data types, but not about how and when they were acquired or inferred. Further, they do not show information such as frequency of purchase. These details are important to meet the goal of improving transparency into company data practices.

\paragraph{Usability of access mechanism needs improvement:}
Our participants had difficulty in comprehending profile data. For example, a participant asked ``What does MOB/branded data mean?'' Another misunderstood the meaning of ``High Confidence.'' To understand the meaning of these and many other entries, we had to read many documents. There is a need to improve comprehensibility of profiles. Accessing profile data was not easy; it was not possible to download profile data for easier analysis. For example, each BlueKai page had only five entries and a user with 99 pages had to click on each page to see its contents. 

\paragraph{Implications for public policy management:}
Users would benefit if companies that create behavioral profiles provide better notice about collection, combining and potential uses of user data. Improving awareness of access mechanisms among users can also help users. At present, there seems to be little awareness, for example, only one out of eight interview participants knew about access mechanisms. Users would benefit if companies get users' consent before combining data from different contexts. To alleviate users' concerns regarding data use, companies could disclose the purposes for which they use profile data. Further, they could specify what inferences they draw and how their prediction models work. From a user's perspective, stating that the company uses proprietary models, for example, ``developed a proprietary algorithm that utilizes a consumer’s name, mailing address and 320 different data points to accurately assign a personality type to 85\% of US adult consumers~\cite{bluekaibook2013},'' may be insufficient. To address user concerns regarding level of detail of profile data, companies could explain the need for such level of detail. Lastly, users would benefit if companies ensure accuracy in profile data and address the issue of accountability for adverse impact arising from errors in profiles. 

\paragraph{Directions for future research:}
We need user studies to further evaluate usability of profile access mechanisms. Results from our study can inform research in related areas such as online behavioral advertising (OBA) company data practices and privacy notices. For example, studies that evaluate whether access matters to users~\cite{leon2013matters} could use realistic behavioral profiles from our study; research on making privacy policies more usable for users~\cite{sadeh2014usablepp} could extract and highlight, from a privacy policy, parts that are of concern to a user. Research on tracking have largely focused on tracking based on online Internet activities~\cite{hoofnagle2012behavioral, krishnamurthy2009privacy, mayer2012third, roesnertracking2012}. However, behavioral profiles contain data from multiple sources including offline sources, and these could be investigated. 


\section{Acknowledgement}
This work was in part supported by the National Science Foundation under grant CNS13-30596 and in part by grants from Google and Samsung. The authors thank Birendra Jha and Alessandro Acquisti for useful suggestions.

\bibliographystyle{abbrv}
\bibliography{whatsinprofile_arxiv}

\begin{thebibliography}{10}

\bibitem{acxiom2014atd}
Acxiom.
\newblock About the data.
\newblock \url{https://aboutthedata.com/}, 2014.
\newblock Accessed: 2014-11-30.

\bibitem{agarwal2013not}
L.~Agarwal, N.~Shrivastava, S.~Jaiswal, and S.~Panjwani.
\newblock Do not embarrass: {R}e-examining user concerns for online tracking
  and advertising.
\newblock In {\em Proceedings of the Symposium on Usable Privacy and Security
  (SOUPS)}. ACM, 2013.

\bibitem{mturk2014}
Amazon.
\newblock Mechanical turk.
\newblock \url{https://www.mturk.com/}, 2014.
\newblock Accessed: 2014-11-30.

\bibitem{bluekaibook2013}
BlueKai.
\newblock Little blue book: {A} buyers guide.
\newblock \url{http://www.bluekai.com/bluebook/}, 2013.
\newblock Accessed: 2014-11-30.

\bibitem{bk2014demographics}
BlueKai.
\newblock Premium demographic data.
\newblock \url{http://www.BlueKai.com/bluebook/BlueKai-little-blue-book.pdf},
  2013.
\newblock Accessed: 2014-08-01.

\bibitem{bluekai2014market}
BlueKai.
\newblock {Audience Data Marketplace}.
\newblock \url{https://www.bluekai.com/audience-data-marketplace.php}, 2014.
\newblock Accessed: 2014-11-30.

\bibitem{bluekai2014registry}
BlueKai.
\newblock {The {B}lue{K}ai Registry}.
\newblock \url{http://bluekai.com/registry/}, 2014.
\newblock Accessed: 2014-11-30.

\bibitem{experian2013ethnic}
{Experian Marketing Services}.
\newblock Ethnic insight.
\newblock
  \url{http://www.experian.com/assets/data-university/brochures/ems-list-services-catalog.pdf},
  2012.
\newblock Accessed: 2014-11-30.

\bibitem{experianmarket2013}
{Experian Marketing Services}.
\newblock List services catalog.
\newblock
  \url{http://www.experian.com/assets/data-university/brochures/ems-list-services-catalog.pdf},
  2012.
\newblock Accessed: 2014-11-30.

\bibitem{ftc2014broker}
{Federal Trade Commission}.
\newblock Data brokers: {A} call for transparency and accountability, May 2014.

\bibitem{gomez2009knowprivacy}
J.~Gomez, T.~Pinnick, and A.~Soltani.
\newblock Know privacy.
\newblock {\em UC Berkeley, School of Information}, June 2009.

\bibitem{google2014adsetting}
Google.
\newblock Google ad settings.
\newblock \url{http://www.google.com/settings/ads}, 2014.
\newblock Accessed: 2014-11-30.

\bibitem{hoofnagle2012behavioral}
C.~J. Hoofnagle, A.~Soltani, N.~Good, and D.~J. Wambach.
\newblock Behavioral advertising: The offer you can't refuse.
\newblock {\em Harvard Law and Policy Review}, 2012.

\bibitem{kelley2011users}
P.~G. Kelley, M.~Benisch, L.~F. Cranor, and N.~Sadeh.
\newblock When are users comfortable sharing locations with advertisers?
\newblock In {\em Proceedings of the SIGCHI Conference on Human Factors in
  Computing Systems}, pages 2449--2452. ACM, 2011.

\bibitem{krishnamurthy2009privacy}
B.~Krishnamurthy and C.~Wills.
\newblock Privacy diffusion on the web: {A} longitudinal perspective.
\newblock In {\em Proceedings of the World Wide Web Conference(WWW)}. ACM,
  2009.

\bibitem{leon2013matters}
P.~G. Leon, B.~Ur, Y.~Wang, M.~Sleeper, R.~Balebako, R.~Shay, L.~Bauer,
  M.~Christodorescu, and L.~F. Cranor.
\newblock What matters to users? {F}actors that affect users' willingness to
  share information with online advertisers.
\newblock In {\em Proceedings of the Symposium on Usable Privacy and Security
  (SOUPS)}. ACM, 2013.

\bibitem{mayer2012third}
J.~R. Mayer and J.~C. Mitchell.
\newblock Third-party web tracking: Policy and technology.
\newblock In {\em Proceedings of the Conference on Security and Privacy}. IEEE,
  2012.

\bibitem{mcdonald2010beliefs}
A.~McDonald and L.~Cranor.
\newblock Beliefs and behaviors: {I}nternet users' understanding of behavioral
  advertising.
\newblock In {\em Proceedings of the Research Conference on Communication,
  Information and Internet Policy}. TPRC, 2010.

\bibitem{microsoft2014optout}
Microsoft.
\newblock Microsoft personalized ad preferences.
\newblock \url{https://choice.microsoft.com/en-US/opt-out}, 2014.
\newblock Accessed: 2014-11-30.

\bibitem{pew2013anonymity}
L.~Rainie, S.~Kiesler, R.~Kang, and M.~Madden.
\newblock Anonymity, privacy, and security online.
\newblock {\em {PEW} Research Center}, September 2013.

\bibitem{roesnertracking2012}
F.~Roesner, T.~Kohno, and D.~Wetherall.
\newblock Detecting and defending against third-party tracking on the web.
\newblock In {\em Proceedings of the Symposium on Network Systems design and
  Implementation (NSDI)}. USENIX, 2012.

\bibitem{rosen2012right}
J.~Rosen.
\newblock The right to be forgotten.
\newblock {\em Stanford law review online}, 64:88, 2012.

\bibitem{sadeh2014usablepp}
N.~Sadeh, A.~Acquisti, T.~D. Breaux, L.~F. Cranor, A.~M. McDonald,
  J.~Reidenberg, N.~A. Smith, F.~Liu, N.~C. Russell, F.~Schaub, S.~Wilson,
  J.~T. Graves, P.~G. Leon, R.~Ramanath, and A.~Rao.
\newblock Towards usable privacy policies: Semi-automatically extracting data
  practices from websites' privacy policies (poster).
\newblock In {\em Proceedings of the Symposium on Usable Privacy and Security
  (SOUPS)}. ACM, 2014.

\bibitem{truste2011oba}
{Truste Research}.
\newblock Consumer research results: {P}rivacy and online behavioral
  advertising.
\newblock July 2011.

\bibitem{turow2009americans}
J.~Turow, J.~King, C.~J. Hoofnagle, A.~Bleakley, and M.~Hennessy.
\newblock Americans reject tailored advertising and three activities that
  enable it.
\newblock {\em Available at SSRN 1478214}, 2009.

\bibitem{ur2012smart}
B.~Ur, P.~G. Leon, L.~F. Cranor, R.~Shay, and Y.~Wang.
\newblock Smart, useful, scary, creepy: {P}erceptions of online behavioral
  advertising.
\newblock In {\em Proceedings of the Symposium on Usable Privacy and Security
  (SOUPS)}. ACM, 2012.

\bibitem{v12postal2014}
V12group.
\newblock Consumer in {U}.{S}. with postal address.
\newblock
  \url{http://listselector.com/docs/learnmore/Consumer%20in%20U.S.%20with%20Postal%20Address.pdf},
  2014.
\newblock Accessed: 2014-11-30.

\bibitem{v122014select}
V12group.
\newblock Consumer selects.
\newblock \url{http://www.v12groupinc.com/consumer-selects/}, 2014.
\newblock Accessed: 2014-07-27.

\bibitem{v122014linkage}
V12group.
\newblock Linkage database: Anonymous consumer recognition.
\newblock
  \url{http://www.v12groupinc.com/wp-content/uploads/2013/03/Linkage-Database-2014-.pdf},
  2014.
\newblock Accessed: 2014-07-27.

\bibitem{v122014pyco}
V12group.
\newblock {PYCO} personality database.
\newblock \url{http://www.v12groupinc.com/data/personality-data/}, 2014.
\newblock Accessed: 2013-11-17.

\bibitem{yahoo2014adinterest}
Yahoo.
\newblock Yahoo ad interests.
\newblock
  \url{https://info.yahoo.com/privacy/us/yahoo/opt_out/targeting/details.html},
  2014.
\newblock Accessed: 2014-11-30.

\bibitem{yahoo2014append}
Yahoo.
\newblock Yahoo appended and matched data.
\newblock \url{https://info.yahoo.com/privacy/us/yahoo/appenddata/}, 2014.
\newblock Accessed: 2014-11-30.

\end{thebibliography}
\bigskip

\appendix
 
\section{Online Survey Questionnaire}
\label{sec:appendix}
\noindent
$[$Consent instructions here$]$\\

\noindent
Important: Please think thoroughly before answering each question. Your precise responses are very important for us. We are not interested in what someone else thinks - we want to know what you think! You may give an incomplete answer or say you do not know\\

\noindent
1) We are interested in understanding how you experience things online. We will start by seeking your views about website advertising. Here, ``website advertising'' refers to ads that are displayed on the web pages that you visit. In a sentence or two, please tell us what you think about website advertising.\\

\noindent
2) What is your age (in years)?\\

\noindent
3) What is your gender?\\
( ) Male	( ) Female   ( ) Decline to answer\\

\noindent
4) Which of the following best describes your primary occupation?\\
$[$List of options here$]$\\
( ) Other (Please specify):: \\
( ) Decline to answer\\

\noindent
5) Which of the following best describes your highest achieved education level?\\
( ) No high school\\
( ) Some high school\\
( ) High school graduate\\
( ) Some college - no degree\\
( ) Associates 2 year degree\\
( ) Bachelors 4 year degree\\
( ) Graduate degree - Masters, PhD, professional, medicine, etc.\\
( ) Decline to answer\\

\noindent
6) Do you have a college degree or work experience in computer science, software development, web development or similar computer-related fields?\\
( ) Yes  ( ) No  ( ) Decline to answer\\

\noindent
7) Advertisers can personalize ads on websites to ensure that the ads are relevant to you. \\
Please indicate how much you agree or disagree with the following statement.\\

\noindent
I like personalization of ads on websites.\\
( ) Strongly disagree	 ( ) Disagree	 ( ) Neither agree nor disagree     ( ) Agree	 ( ) Strongly agree\\

\noindent
Advertisers collect data about you in order to personalize ads. Advertisers may create profiles about you using the collected data.\\

\noindent
The following is an image of a profile that shows the different types of data that advertisers collect about users like you. Please look through the entire image at your own pace, and then answer the following questions.\\

\noindent
$[$Sample profile (Fig. 5) here$]$\\

\noindent
8) Please select from the list below at least two items that appear in the sample profile.\\
$[$ $]$ Male \\
$[$ $]$ Credit Card Interest Score 8-9\%\\
$[$ $]$ Offline CPG Purchasers $>$ Charmin Ultra Strong\\
$[$ $]$ Personal Health (Values 70-90\%)\\
$[$ $]$ Interest in Religion Code (Value Tiers 1-3)\\
$[$ $]$ Household Income (HHI) $>$ Income Range \$75,000 - \$99,000\\

\noindent
$[$Randomize Q9 - Q13$]$\\
We will ask you some questions to understand your reaction to the profile you just saw. It is important that you have looked at the different types of data in the profile before continuing. Please click next when you are ready.\\

\noindent
9) Please describe your reaction to the profile.				$[$Fig. 5 shown here$]$\\
Indicate how much you agree or disagree with the following statement.\\
I am concerned because I believe that the profile contains sensitive data.\\
( ) Strongly disagree	 ( ) Disagree	 ( ) Neither agree nor disagree     ( ) Agree	 ( ) Strongly agree\\

\noindent
10) Please describe your reaction to the profile.				$[$Fig. 5 shown here$]$\\
Indicate how much you agree or disagree with the following statement.\\
I am concerned by the amount of data in the profile.\\
( ) Strongly disagree	 ( ) Disagree	 ( ) Neither agree nor disagree     ( ) Agree	 ( ) Strongly agree\\

\noindent
11) Please describe your reaction to the profile.				$[$Fig. 5 shown here$]$\\
Indicate how much you agree or disagree with the following statement.\\
I am concerned because my data from multiple sources (e.g. online activities, in-store, other companies) is being combined.\\
( ) Strongly disagree	 ( ) Disagree	 ( ) Neither agree nor disagree     ( ) Agree	 ( ) Strongly agree\\

\noindent
12) Please describe your reaction to the profile.				$[$Fig. 5 shown here$]$\\
Indicate how much you agree or disagree with the following statement.\\
I am concerned by the level of detail (e.g. specific information, not just broad categories) in the profile.\\
( ) Strongly disagree	 ( ) Disagree	 ( ) Neither agree nor disagree     ( ) Agree	 ( ) Strongly agree\\

\noindent
13) Please describe your reaction to the profile. 				$[$Fig. 5 shown here$]$\\
Indicate how much you agree or disagree with the following statement.\\
I am concerned about how my data may be used.\\
( ) Strongly disagree	 ( ) Disagree	 ( ) Neither agree nor disagree     ( ) Agree	 ( ) Strongly agree\\

\noindent
14) Please explain if you have other concerns about the profile.\\

\noindent
You are almost done. \\

\noindent
We will ask you how you feel about personalized ads after seeing the profile. We will also give you a chance to look at your own profile. Please note that looking at your own profile is optional.\\

\noindent
15) Please indicate how much you agree or disagree with the following statement.\\
After seeing the types of data collected for personalization, my liking for personalized ads on websites has decreased.\\
( ) Strongly disagree	 ( ) Disagree	 ( ) Neither agree nor disagree     ( ) Agree	 ( ) Strongly agree\\

\noindent
You looked at a sample profile. Would you like to look at your own profile and learn what data advertisers have about you? \\

\noindent
Please note that this is optional. Your payment and bonus will not be affected if you choose to skip looking at your own profile. However, what you learn may be beneficial to you.\\

\noindent
16) Would you like to look at your own profile?\\
( ) Yes 	( ) No\\

\noindent
Thank you for choosing to look at your own profile. We believe it will be beneficial to you and us.\\

\noindent
Please copy and paste the following website link in a new tab or window to access your own profile. You should see a profile similar in appearance to the sample profile. \\
http://bluekai.com/registry/\\

\noindent
Please note that the profile may not display properly if you have disabled browser cookies. You can try from a different browser if you have more than one browser installed.\\

\noindent
If you are not able to access your profile using the above link, you can alternatively try the following websites.\\
\url{https://aboutthedata.com/} (scroll to the bottom of the page to click on ``See and Edit Marketing Data about You.'')\\
\url{http://www.google.com/settings/ads}\\
\url{http://info.yahoo.com/privacy/us/yahoo/opt_out/targeting/}\\

\noindent
17) Please tell us briefly what you found in your own profile and how you feel about it (optional but helpful for our research).\\

\noindent
18) Do you have any further comments?\\

Thank you for taking our survey. Your response is important to us. Below is your confirmation code. You must retain this code to be paid -- it is recommended that you store your code in a safe place (either by writing it down, or by printing this page). \\

%
\end{document}